\documentstyle[here,epsfig]{mn2e}
\def\simlt{\mathrel{\rlap{\lower 3pt\hbox{$\sim$}}\raise 2.0pt\hbox{$<$}}}
\def\simgt{\mathrel{\rlap{\lower 3pt\hbox{$\sim$}} \raise 2.0pt\hbox{$>$}}}

\def\gtsima{$\; \buildrel > \over \sim \;$}
\def\ltsima{$\; \buildrel < \over \sim \;$}
\def\gtrsim{\lower.5ex\hbox{\gtsima}}
\def\lesssim{\lower.5ex\hbox{\ltsima}}
\def\url#1{{\ttfamily\def\/{/\diskretionary{}{}{}}#1}}
\begin{document}

\newcommand{\q}{\begin{equation}}
\newcommand{\qa}{\begin{eqnarray}}
\newcommand{\qs}{\begin{eqnarray*}}
\newcommand{\nq}{\end{equation}}
\newcommand{\nqa}{\end{eqnarray}}
\newcommand{\nqs}{\end{eqnarray*}}
\newcommand{\ud}{\mathrm{d}}

\title[IMBHs in dwarf galaxies] 
{Intermediate-mass black holes in dwarf galaxies: the case of Holmberg II}
\author[M. Mapelli] 
{M. Mapelli$^{1,2}$ \\
$^{1}$SISSA, International School for Advanced Studies, Via Beirut 4, I-34014, Trieste, Italy; {\tt mapelli@sissa.it}\\
$^{2}$ Institute for Theoretical Physics, University of Z\"urich, Winterthurerstrasse 190, CH-8057, Z\"urich, Switzerland\\}

\maketitle \vspace {7cm }

\begin{abstract}
In order to constrain the density of intermediate-mass black holes  (IMBHs) in galaxies, we run smoothed particle hydrodynamics (SPH) simulations of a gas-rich disc dwarf galaxy, where different halo and disc populations of IMBHs are embedded. 
IMBHs, when passing through dense gas regions, can accrete gas and switch on as X-ray sources. 
We derive the luminosity distribution of simulated IMBHs, by assuming that they accrete at the Bondi-Hoyle rate. 
The X-ray distribution of simulated IMBHs has been compared with that of observed sources in the dwarf galaxy Holmberg II, chosen for its richness in gas, its small mass (compared to spiral galaxies), and the accuracy of the available X-ray measurements. Holmberg II also hosts one of the strongest IMBH candidates. From this comparison, we find that the density parameter of disc (halo) IMBHs must be $\Omega{}_\bullet{}\lesssim{}10^{-5}\,{}\Omega{}_b$ ($\Omega{}_\bullet{}\lesssim{}10^{-2}\,{}\Omega{}_b$, where $\Omega{}_b$ is the density parameter of baryons),  for a radiative efficiency $10^{-3}$ and an IMBH mass of $10^4M_\odot{}$. These constraints imply that a dwarf galaxy like Holmberg II cannot host more than 1 (1000) disc (halo) $10^4M_\odot{}$ IMBHs.

\end{abstract}

\section{Introduction}
The existence of intermediate-mass black holes (IMBHs), i.e. black holes (BHs) with mass from $\sim{}20$ to $10^5M_\odot{}$, is still controversial (see van der Marel 2004 for a review). Various IMBH formation processes have been proposed: (i)  collapse of massive ($\gtrsim{}260\,{}M_\odot{}$) metal free stars (Heger \& Woosley 2002); (ii)  runaway collision of stars in young clusters (Portegies Zwart \& McMillan 2002); (iii) repeated mergers of stellar mass BHs in globular clusters (Miller \& Hamilton 2002), and (iv) direct collapse of dense, low angular momentum gas (Haehnelt \& Rees 1993; Umemura, Loeb \& Turner 1993; Loeb \& Rasio 1994; Eisenstein \& Loeb 1995; Bromm \& Loeb 2003), driven by turbulence (Eisenstein \& Loeb 1995) or gravitational instabilities (Koushiappas, Bullock \& Dekel 2004; Begelman, Volonteri \& Rees 2006; Lodato \& Natarajan 2006).

 Recent spectroscopic and photometric measurements suggest the presence of IMBHs in globular clusters (Gebhardt, Rich \& Ho 2002, 2005; Gerssen et al. 2002; van den Bosch et al. 2006). Furthermore, IMBHs accreting dense gas (Krolik 2004; Mii \& Totani 2005) or matter from   companion stars (Hopman, Portegies Zwart \& Alexander 2004; Kalogera et al. 2004; Portegies Zwart, Dewi \& Maccarone 2004; Hopman \& Portegies Zwart 2005; Patruno et al. 2005; Baumgardt et al. 2006) have been invoked to explain the presence of ultra-luminous X-ray sources (ULXs), i.e. X-ray sources with luminosity higher than 10$^{39}$ erg s$^{-1}$ (see Colbert \& Miller 2005 for a review). 

In general, IMBHs accreting  gas in molecular clouds are expected to switch on as X-ray sources (either ultra-luminous or not).
Mapelli, Ferrara \& Rea (2006; hereafter MFR) calculated, via dedicated N-body simulations, the number of X-ray sources which are expected to  originate from IMBHs accreting gas in the Milky Way.   
They compared the number of simulated sources with the number of unidentified X-ray sources observed in our Galaxy, in order to find an upper limit on the number of IMBHs. The results of MFR depend on the adopted accretion model and on the spatial IMBH distribution. In a conservative case, i.e. assuming that the accretion mechanism on IMBHs is an advection dominated accretion flow (ADAF) with radiative efficiency $\eta{}=10^{-3}$, the density of 10$^4\,{}M_\odot{}$ IMBHs in our Galaxy must be  $\lesssim{}10^{-1}\Omega{}_b$ (where $\Omega{}_b=0.042$ is the density of baryons in terms of the critical density of the Universe, Spergel et al. 2006), if the IMBHs are distributed according to a Navarro Frenk \& White (1996; NFW) profile. Diemand, Madau \& Moore (2005; DMM) recently found that objects formed in high $\sigma{}$-fluctuations follow a more concentrated distribution. Adopting the DMM profile, the upper limit on the IMBH density becomes  $\lesssim{}10^{-2}\Omega{}_b$.

In this paper we adopt the same technique developed in MFR (with some substantial improvements, such as the treatment of the gas and of the dark matter halo), in order to put constraints on the IMBH population of Holmberg II (=UGC~4305=DDO~50; hereafter HoII), a dwarf irregular galaxy belonging to the M81 group (Karachentsev et al. 2002).  We chose a dwarf galaxy, and in particular HoII, for various reasons. First of all, there are many papers aimed to constrain the number of IMBHs in the Milky Way,  from a dynamical point of view (Carr \& Sakellariadou 1999; see also Lacey \& Ostriker 1985; Wasserman \& Salpeter 1994; Murali, Arras \& Wasserman 2000 and reference here), from the comparison with X-ray sources (Mii \& Totani 2005; MFR) and from the study of globular cluster formation and disruption (Ostriker, Binney \& Saha 1989; Moore 1993; Klessen \& Burkert 1996; Arras \& Wasserman 1999; Murali et al. 2000); whereas there are no papers dedicated to constrain IMBHs in dwarf galaxies. Is the density of IMBHs comparable in dwarf and normal galaxies? A dependency of the density of IMBHs on the mass of the host galaxy could have important implications on the theoretical models of IMBH formation.

Furthermore, dwarf galaxies are much less massive than the Milky Way, allowing us to reach  higher mass and spatial resolution in our simulations ($\sim{}10^2\,{}M_\odot{}$ and $\lesssim{}1$ pc, respectively, in the case of HoII, versus $5\times{}10^4\,{}M_\odot{}$ and $\sim{}100$ pc for the Milky Way; see MFR).

Among the possible candidates, HoII is particularly suitable for our analysis, because it is a very gas-rich dwarf galaxy, about one tenth of its kinematic mass being represented by HI (Puche et al. 1992; Stewart et al. 2000; Bureau \& Carignan 2002). Therefore, the probability that an IMBH passes through a dense gas region is much higher in HoII than in the Milky Way. 
In addition, the rotation curve (and, consequently, the dynamical mass) and the other dynamical properties of HoII have been accurately measured in the last years (Bureau \& Carignan 2002 and references there).
Finally, the X-ray sources detected in HoII have been reasonably well studied (Kerp, Walter \& Brinks 2002), especially X-1, which is one of the most powerful observed ULXs (Dewangan et al. 2004) and one of the strongest IMBH candidates. By the way, the presence of this ULX ensures  that, if IMBHs exist and are connected with ULXs, HoII is one of the best places where we can find them.  


More details about the observed properties of HoII are given in Section 2. In Section 3 we describe our simulations. The results have been presented and compared with observations in Section 4 (for IMBHs accreting molecular hydrogen) and Section 5 (atomic hydrogen). Section 6 is a summary of  the main findings.

\section{Observational properties of HoII}
HoII is a very gas-rich (Puche et al. 1992; Stewart et al. 2000; Bureau \& Carignan 2002) dwarf irregular, belonging to the M81 group (Karachentsev et al. 2002). 
Its kinematic total mass, derived from HI measurements (Bureau \& Carignan 2002), is $6.3\times{}10^{9}M_\odot{}$, and it is mainly dark ($\sim{}80$ per cent). The mass in HI is 6.44$\times{}10^{8}M_\odot{}$, indicating that there is more luminous mass in gas than in stars. The circular velocity, $\sim{}$36 km s$^{-1}$ (see Bureau \& Carignan 2002 and Fig.~\ref{fig:fig1}), implies that this galaxy rotates quite slowly.
There are no signs of interaction between HoII and close-by members of the M81 group (Kar 52 and UGC~4483), and HoII is only slightly affected by ram pressure (in form of a faint HI tail; Bureau \& Carignan 2002). So it makes sense to simulate HoII as an isolated galaxy.

Kerp et al. (2002), by using ROSAT data, detect 31 X-ray sources (with luminosity $L_X\gtrsim{}10^{37}$ erg s$^{-1}$) located in HoII. To avoid contamination from background sources, they consider only those sources which have been identified in at least one additional frequency range. By adopting this selection criterion, Kerp et al. (2002) are left with 13 sources, which should belong to HoII. Some of them can be associated with either X-ray binaries or supernova remnants (Kerp et al. 2002), but this identification is not certain. Then,  in this paper, we will make the very conservative assumption that all these  13 sources are unidentified and could be, in principle, IMBHs.


The brightest among these X-ray sources, HoII X-1, is one of the most powerful ULXs observed up to now, having a X-ray luminosity $L_X\sim{}2\times{}10^{40}$ erg s$^{-1}$ in the brightest state (Dewangan et al. 2004). An ionized nebula has been detected around HoII X-1 (Pakull \& Mirioni 2003; Kaaret, Ward \& Zezas 2004; Lehmann et al. 2005), and important radio emission has been shown to be connected with it (Miller, Mushotzky \& Neff 2005). The spectrum of HoII X-1 can be fitted combining an absorbed power law and a thermal component (thermal plasma or multi-colour blackbody), which indicates relatively low temperatures ($\sim{}0.14-0.22$ keV; Lehmann et al. 2005). For all these reasons, the hypothesis that HoII X-1 is a beamed X-ray binary is disfavoured, and this source is considered one of the strongest IMBH candidates (Miller, Fabian \& Miller 2004).

\section{Numerical simulations}
As in MFR, the simulations have been carried out using the parallel N-body code GADGET-2 (Springel 2005). The simulations were performed using 8 nodes of the 128-processor cluster {\it Avogadro} at the Cilea ({\url http://www.cilea.it}).  We simulated a N-body model of HoII, in which we embed a halo population of IMBHs.

\subsection{HoII model}
To simulate a model of HoII, we combined three different components: a NFW halo, a stellar exponential disc and a uniform molecular gas disc (or an exponential atomic gas disc).

Halo, disc and gas particle velocities are generated using the Gaussian approximation (Hernquist 1993), as described in MFR.
In the following, we describe the details of our model for halo, stellar disc and gas. All the main adopted parameters of this model are listed in Table 1.

\subsubsection{Halo}
\begin{figure}
\center{{
\epsfig{figure=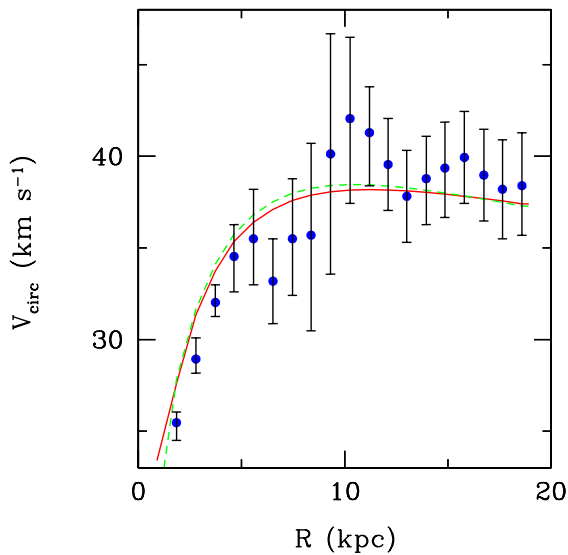,height=8cm}
}}
\caption{\label{fig:fig1} 
Rotation curve of HoII. The solid (dashed) line refers to the average circular velocity of disc star particles assuming a halo concentration $c=3$ and a uniform disc of molecular gas (exponential disc of atomic hydrogen). Data points are from Bureau \& Carignan (2002), and refer to HI measurements.
} 
\end{figure}

The parameters of the NFW halo have been derived as described in MFR. In particular, the density distribution of the halo is (NFW; Moore et al. 1999):
\q\label{eq:eq1}
\rho_h(r)=\frac{\rho_s}{(r/r_s)^\gamma{}\,{}[1+(r/r_s)^\alpha{}]^{(\beta{}-\gamma{})/\alpha{}}},
\nq
where we choose $(\alpha{},\beta{}, \gamma{})=(1,3,1)$, and
$\rho{}_s=\rho_{crit}\,{}\delta_c$, $\rho_{crit}$ being the critical
density of the Universe and
\begin{equation}\label{eq:eq2}
\delta_c=\frac{200}{3}\frac{c^3}{\ln{(1+c)}-[c/(1+c)]},  
\end{equation}
where $c$ is the concentration parameter and $r_s$ is the halo scale radius, defined by
$r_s=R_{200}/c$. $R_{200}$ is the radius encompassing a mean overdensity of 200 with respect to the 
background density of the Universe.
  $R_{200}$ can be calculated as $R_{200}=V_{200}/[10\,{}H(z_{vir})]$, where $V_{200}$ is the circular velocity at the
virial radius and $H(z_{vir})$ is the Hubble parameter at the virialization redshift $z_{vir}$. For HoII, we adopt $V_{200}=36$ km s$^{-1}$ (Bureau \& Carignan 2002) and $z_{vir}=1.8$, obtaining $R_{200}\simeq{}20$ kpc, in agreement with HI observations (Bureau \& Carignan 2002). 
In order to reproduce the observed rotation curve (Fig.~\ref{fig:fig1}), we have to assume $c\leq{}3$, which is a very low concentration. Cosmological simulations (Bullock et al. 2001) show that dwarf galaxies should have significantly higher concentration ($c\gtrsim{}10$). However such high concentrations cannot account for the observed rotation curve, indicating that the NFW profile fails to predict the properties of dwarf irregular galaxies.
For the problem of fitting dwarf galaxy rotation curves by using a NFW profile see Navarro (1998) and Mo \& Mao (2000).

Adopting the parameters listed above, the total mass within $R_{200}$ is $M_{200}=5.9\times{}10^9M_\odot{}$, in nice agreement with observations ($6.3\times{}10^9M_\odot{}$, Bureau \& Carignan 2002). The corresponding mass of dark matter within $R_{200}$ is $M_{\rm DM}=M_{200}-M_d-M_{gas}$, where $M_d$ is the stellar disc mass, and $M_{gas}$ is the gas mass ($M_{gas}$ is equal to the mass in molecular gas, $M_{mol}$, or to the mass in atomic gas,  $M_{\rm HI}$, depending on whether we simulate molecular or atomic gas, see Section 3.1.3).

\subsubsection{Disc}
The stellar disc profile is (Hernquist 1993):
\q\label{eq:eq3}
\rho_d(R,z)=\frac{M_d}{4\pi R_d^2\,{}z_0}\,{}e^{-R/R_d}\,{}\textrm{sech}^2(z/z_0),
\nq
where $M_d=4\times{}10^8\,{}M_\odot{}$ is the disc mass, $R_d=2$ kpc the disc scale length and $z_0=0.1\,{}R_d$ the disc scale height. These adopted parameters are in good agreement with observations (Bureau \& Carignan 2002) and correspond to a spin parameter $\lambda{}\simeq{}0.07$, indicating a dynamically stable disc (Mo, Mao \& White 1998). 

\subsubsection{Gas component}
The gas in HoII is expected to be partially molecular and partially atomic. We are particularly interested in the molecular gas, because it is expected to be more efficiently accreted by IMBHs (being much denser than the atomic hydrogen). However,  we  want to check the possibility that also IMBHs accreting atomic hydrogen are observable as X-ray sources (see MFR for the Milky Way). To simulate these two gas components simultaneously is beyond the purpose of this paper. Thus, we made two different sets of runs (see Section 3.3), part of them including only molecular gas and the remaining only atomic gas. In runs with molecular gas the mass of gas which is expected to be in atomic form is attributed to dark matter particles.

We chose the profile of the gas component according to whether the gas is assumed to be molecular or atomic.
The molecular gas has been represented as a uniform disc with scale length $R_{mol}=500$ pc, scale height $z_{mol}=60$ pc (Taylor et al. 1999) and mass $M_{mol}=2\times{}10^8M_\odot{}$. In fact, HoII has a total HI mass $6.44\times{}10^8M_\odot{}$ (Bureau \& Carignan 2002) and we know that the average fraction of H$_2$ versus HI in dwarf irregular galaxies is $\sim{}0.3$ (Leroy et al. 2005).

 The adopted equation of state is polytropic:
\q\label{eq:eq4}
P_{gas}=\kappa{}\,{}\rho{}_{gas}^\Gamma{},
\nq
where $P_{gas}$ and $\rho{}_{gas}$ are the gas pressure and density, $\Gamma{}$ is equal to 5/3 (which avoids excessive fragmentation, ensuring the stability of the disc, Escala et al. 2005) and $\kappa{}$ is a parameter corresponding to the entropy of the gas. $\kappa{}$ is related to the gas temperature by the following relation.
\q\label{eq:eq5}
T_{gas}\simeq{}60\,{}\textrm{K}\,{}\left(\frac{\kappa{}}{9\times{}10^{23}\textrm{g}^{-(\Gamma{}-1)}\,{}\textrm{cm}^4\textrm{s}^{-2}}\right)\,{}\left(\frac{n_{gas}}{100\textrm{cm}^{-3}}\right)^{(\Gamma{}-1)}\,{}\left(\frac{\mu{}}{2}\right)^{\Gamma{}},
\nq
where $n_{gas}$ is the gas density and $\mu{}$ the mean molecular weight ($\mu{}=2$ for molecular gas). 
 We consider values of $\kappa{}$ ranging from $9\times{}10^{23}\textrm{g}^{-2/3}\,{}\textrm{cm}^4\textrm{s}^{-2}$ (corresponding to a gas temperature $T_{gas}=60$ K, suitable for molecular gas; see Spitzer 1998) to $1.5\times{}10^{25}\textrm{g}^{-2/3}\,{}\textrm{cm}^4\textrm{s}^{-2}$ (corresponding to a gas temperature $T_{gas}=1000$ K, as an upper limit). Discs with a low $\kappa{}$ tend to become clumpy before hotter discs (Escala et al. 2005), and have a smaller Jeans mass. On the other hand, our choice of  $\Gamma{}=5/3$ is very conservative, and balances the effect of $\kappa{}$, avoiding extreme fragmentation of the disc.

Instead, when we simulate atomic gas, we assume an exponential distribution (the same as equation~(\ref{eq:eq3})), according to the observed neutral hydrogen exponential profile in disc galaxies (Lockman 2002). The total mass of atomic hydrogen is assumed to be $M_{\rm HI}=6.44\times{}10^8\,{}M_\odot{}$ (Bureau \& Carignan 2002), the scale length of the disc is $R_{\rm HI}=5$ kpc, and the scale height $z_{\rm HI}=600$ pc (Table 1). The equation of state assumed for the neutral atomic gas is the same as that for the molecular gas (with $\mu{}=1.3$, $T_{gas}=60$ K and $n_{gas}=1\textrm{ cm}^{-3}$). 

In the simulations with atomic exponential gas, 
we adopt the same values of $c$, $R_{200}$, $M_{200}$, $M_d$, $R_d$ and $z_0$ (Table 1) that we used for the runs with molecular gas. 
In fact, the rotation curve is nearly the same for runs with molecular and with atomic hydrogen (Fig.~\ref{fig:fig1}), because the gas component (about one tenth of the dark matter mass)
 has a minor impact on the dynamics.
 Thus, the only differences between runs with molecular and with atomic hydrogen are the treatment of the gas component and  the total mass in dark matter\footnote{When we run simulations with molecular gas but without atomic hydrogen, we have to redistribute among dark matter particles the mass which is expected to be in form of atomic hydrogen (see Section 3.3).}. This choice  makes  easier the comparison between the two cases, since the main characteristics of the galaxy model are unchanged.

\begin{table}
\begin{center}
\caption{Initial parameters for the HoII model.
}
\begin{tabular}{ll}
\hline
\vspace{0.1cm}
$c$ & 3\\
$V_{200}$ & 36 km s$^{-1}$\\
$R_{200}$ & 19.6 kpc \\
$M_{200}$ & 5.9$\times{}10^{9}M_{\odot{}}$\\
$\lambda{}$ & 0.07\\
$M_d$ & 4$\times{}10^{8}M_\odot{}$\\
$R_d$ & 2 kpc\\
$z_0$ & 0.1 $R_d$\\
\vspace{0.05cm}
$M_{mol}$ & 2$\times{}10^{8}M_{\odot{}}$\\
$R_{mol}$ & 500 pc \\
$z_{mol}$ & 60 pc\\
\vspace{0.05cm}
$M_{\rm HI}$ & 6.44$\times{}10^{8}M_{\odot{}}$\\
$R_{\rm HI}$ & 5 kpc \\
$z_{\rm HI}$ & 600 pc\\
\hline
\end{tabular}
\end{center}
\label{tab_1}
\end{table}
\subsection{Intermediate-mass black holes}
The number of IMBHs hosted in galaxies is completely unknown. If we assume that they formed at high redshift, then an indicative estimate of the IMBH number has been derived in Volonteri, Haardt \& Madau (2003). As in MFR, we use the formalism of Volonteri et al. (2003) to generate the initial conditions of our simulations. In particular, we estimate the number of IMBHs residing in a galaxy ($N_\bullet{}$) as
\q\label{eq:numBH}
N_\bullet{}\sim{}\frac{\Omega{}_\bullet{}}{\Omega{}_b}\,{}\frac{M_b}{m_\bullet{}},
\nq
where $\Omega{}_\bullet{}$ and $\Omega{}_b$(=0.042; Spergel et al. 2006) are the current density of IMBHs and baryons, respectively, in terms of the critical density of the Universe; $M_b$ is the mass in baryons of the host galaxy, and $m_\bullet{}$ is the current average IMBH mass. For HoII $M_b\sim{}10^9\,{}M_\odot{}$, so that equation~(\ref{eq:numBH}) becomes $N_\bullet{}\sim{}100\,{}(\Omega{}_\bullet{}/10^{-3}\Omega_b)\,{}(10^4\,{}M_\odot{}/m_\bullet{})$. According to MFR, the case with $\Omega{}_\bullet{}=10^{-3}\Omega_b$ corresponds to assume that IMBHs formed in 3 $\sigma{}$ fluctuations at redshift $z\sim{}20-25$. 

This estimate is affected by a number of uncertainties, and we adopt it only because it is a straightforward formalism to express the number of IMBHs in terms of the critical density of the Universe. However, the results of our simulations have more general validity, and can be applied also to IMBHs formed via different mechanisms (e.g. runaway collapse).

The spatial and velocity distribution of IMBHs are another severe problem. If IMBHs formed in high redshift minihalos, as remnants of metal free stars (Volonteri et al. 2003), they probably constitute a halo population in current galaxies. For this reason, Mii \& Totani (2005) chose a NFW profile to represent the distribution of IMBHs in their model. However, White \& Springel (2000) suggested that remnants of population III stars should be more concentrated inside present-day halos than younger objects. DMM confirmed this idea and showed that the current spatial distribution of objects formed in high-$\sigma{}$ fluctuations depends only on the rarity of the peak in which they are born. DMM rewrote the NFW profile, accounting for this correction, as
\q\label{eq:DMM}
\rho{}_{\bullet{}}(r)=\frac{\rho_s}{(r/r_\nu)^\gamma{}\,{}[1+(r/r_\nu{})^\alpha{}]^{(\beta{}_\nu{}-\gamma{})/\alpha{}}},
\nq
where $\alpha{}$ and $\gamma{}$ are the same as defined in the previous section;  $r_\nu\equiv{}r_s/f_\nu{}$ is the scale radius for objects formed in a  $\nu{}\,{}\sigma{}$ fluctuation (with $f_\nu{}=\exp{(\nu{}/2)}$), and $\beta{}_\nu{}=3+0.26\,{}\nu{}^{1.6}$. Hereafter, we will refer to equation~(\ref{eq:DMM}) as the DMM profile. This profile is likely to track the distribution of IMBHs, if they are a halo population (MFR).

On the other hand, IMBHs could also be a disc population, either if they form by runaway collapse in young clusters, or if they are remnants of population III stars born in the disc. The latter scenario is not so unrealistic, if a
 part of the metal free gas (i.e. $Z\lesssim{}10^{-4}$, Bromm et al. 2001; Schneider et al. 2002) remains unpolluted even at low redshift, forming population III stars in the disc of galaxies (Krolik 2004). For example, Jimenez \& Haiman (2006) argued that a percentage of metal free stars as high as 10-50 per cent of the total stellar population is required at redshift $z\sim{}3-4$ to explain some puzzling observations (e.g. the significant ultraviolet emission from Lyman Break Galaxies at wavelengths shorter than 912 \AA{}). If IMBHs are a disc population, they are so massive that dynamical friction exerted by gas is expected to pull them toward the galaxy center in a short time-scale ($\sim{}3\times{}10^8$ yr, for $10^4\,{}M_\odot{}$ IMBHs, if the average density of gas in the stellar disc is  $\sim{}$1 cm$^{-3}$; Binney \& Tremaine 1987). Then, we expect that their distribution is more concentrated than that of other disc populations.

We will examine both these cases, i.e. IMBHs distributed in the halo and in the disc. For the halo IMBHs, we adopt a DMM profile, as described in MFR. In MFR we also discussed a NFW distribution; but this distribution is quite unlikely for halo IMBHs (White \& Springel 2000; DMM). Then, we decided not to consider this case in the present paper. Furthermore, MFR showed that the constraints for NFW IMBHs have the same trend of those derived for DMM IMBHs, but are a factor of 10 weaker. 
For disc IMBHs, we assume an exponential distribution, similar to that described for stars, but with a cut-off radius  $R_c=1$ kpc, a scale radius $R_{d,\,{}BH}=500$ pc and a scale height $R_{d,\,{}BH}=10$ pc, accounting for dynamical friction.

\subsection{Description of runs}
\begin{table*}
\begin{center}
\caption{Initial parameters for IMBHs and gas distribution.
}
\begin{tabular}{llllll}
\hline
\vspace{0.1cm}
Run & IMBH mass ($M_\odot{}$) & Number of IMBHs & $\Omega_{\bullet{}}/\Omega_b$ & IMBH profile & Gas profile\\
\hline
disc4\_{}10       & 10$^4$ & 10   & 10$^{-4}$ & exponential disc & uniform disc\\
disc4\_{}100      & 10$^4$ & 100  & 10$^{-3}$ & exponential disc & uniform disc\\
disc5\_{}10       & 10$^5$ & 10   & 10$^{-3}$ & exponential disc & uniform disc\\
\vspace{0.1cm}
K\_{}disc4\_{}100 & 10$^4$ & 100  & 10$^{-3}$ &  exponential disc & uniform disc\\
DMM4\_{}100       & 10$^4$ & 100  & 10$^{-3}$ & DMM & uniform disc\\
DMM4\_{}1000      & 10$^4$ & 1000 & 10$^{-2}$ & DMM & uniform disc\\
DMM5\_{}100       & 10$^5$ & 100  & 10$^{-2}$ & DMM & uniform disc\\
\vspace{0.1cm}
DMM5\_{}1000      & 10$^5$ & 1000 & 10$^{-1}$ & DMM & uniform disc\\
HI\_{}disc4\_{}10 & 10$^4$ & 10   & 10$^{-4}$ & exponential disc & exponential disc \\
HI\_{}DMM4\_{}100 & 10$^4$ & 100  & 10$^{-3}$ & DMM & exponential disc \\
HI\_{}DMM5\_{}100 & 10$^5$ & 100  & 10$^{-2}$ & DMM & exponential disc \\
\hline
\end{tabular}
\end{center}
\label{tab_2}
\end{table*}
The simulated dwarf galaxy is composed by:
\begin{itemize}
\item[-] 10$^6$ halo particles inside $R_{200}$, each one of $5.3\times{}10^3\,{}M_\odot{}$ or  $4.9\times{}10^3\,{}M_\odot{}$ (in the case we simulate a uniform disc of H$_2$ or an exponential disc of atomic gas, respectively);
\item[-] 10$^6$ disc particles, each one of $4\times{}10^2\,{}M_\odot$;
\item[-] 10$^6$ smoothed particle hydrodynamics (SPH) gas particles,  each one of $2\times{}10^2\,{}M_\odot{}$ or  $6.4\times{}10^2\,{}M_\odot{}$ (in the case we simulate a uniform disc of H$_2$ or an exponential disc of atomic gas, respectively).
\end{itemize} 
The gravitational softening lengths are $l_{soft}=6$, 0.2 and 0.1~pc for halo, disc and gas particles, respectively. The average initial SPH length\footnote{Bate \& Burkert (1997) showed that, if $l_{\rm SPH}\sim{}l_{J}$, where $l_{J}$ is the Jeans length, and if $l_{\rm SPH}\gg{}l_{soft}$ or $l_{\rm SPH}\ll{}l_{soft}$, some unphysical effects can be induced. In our simulations we always have $l_{J}\sim{}10-30\textrm{ pc}>l_{\rm SPH}$. Thus, even if the balance between pressure and gravity at distances smaller than $\sim{}$2 pc can be affected by our choice of $l_{\rm SPH}$ and $l_{soft}$, the Jeans mass is resolved by the simulations. Furthermore, we compared our simulations with some test runs with $l_{soft}=6$, 2 and 2 pc (=$l_{\rm SPH}$) for halo, disc and gas particles, respectively, and we found negligible differences.} is $l_{\rm SPH}\sim{}2$ pc.

It is worth noting that (i) we simulated a 'living' halo, rather than the rigid halo adopted by MFR; (ii) we make an SPH treatment of gas particles, whereas MFR used a Monte Carlo technique to account for molecular and atomic gas in their merely N-body simulations.

In this galaxy model we followed the dynamics of various populations of IMBHs. The mass of each IMBH is 10$^4$ or 10$^5$ $M_\odot{}$ depending on the run. We also made some checks with higher mass BHs (10$^{6-7}$ $M_\odot{}$). We did not consider lower mass BHs, because of our mass resolution. The number of IMBHs per simulation is 10$-$1000 (corresponding to $\Omega{}_\bullet{}=10^{-4}-10^{-2}\,{}\Omega{}_b$). IMBHs were distributed according to a DMM profile or an exponential disc. The softening of IMBH particles is as small as possible (0.002 pc).

We made more than 30 different runs; but, for simplicity, in Table~2 we list only the most significant among them. In eight of the runs reported in Table~2 (disc4\_{}10, disc4\_{}100, disc5\_{}10, K\_{}disc4\_{}100, DMM4\_{}100, DMM4\_{}1000, DMM5\_{}100 and DMM5\_{}1000) we simulated a uniform disc of molecular gas; while in the remaining three (HI\_{}disc4\_{}10, HI\_{}DMM4\_{}100 and HI\_{}DMM5\_{}100, i.e. the runs labelled as 'HI') we simulated an exponential disc of atomic hydrogen. 

IMBHs are assumed to be a disc population in five of these runs (disc4\_{}10, disc4\_{}100, disc5\_{}10, K\_{}disc4\_{}100 and HI\_{}disc4{}\_{}10, i.e. the runs labelled as 'disc') and a halo population, following a DMM profile, in the other six runs (DMM4\_{}100, DMM4\_{}1000, DMM5\_{}100, DMM5\_{}1000, HI\_{}DMM4\_{}100 and HI\_{}DMM5\_{}100, i.e. the runs labelled as 'DMM').
Finally, those runs whose label ends in 4\_{}$n$ (5\_{}$n$) host a number $n$ (with $n$=10, 100 or 1000) of 10$^4$ $M_\odot{}$ (10$^5$ $M_\odot{}$) IMBHs.

The gas has $\kappa{}=9\times{}10^{23}$ g$^{-2/3}$ cm$^4$ s$^{-2}$ (see equation~\ref{eq:eq4}) in all the cases reported in Table~2, except for the run K\_{}disc4\_{}100, where $\kappa{}=1.5\times{}10^{25}$ g$^{-2/3}$ cm$^4$ s$^{-2}$.

Each run was stopped after 6 Myr. A longer time would require an excessive amount of computational time (i.e. more than 2 weeks on 16 3-GHz CPUs per run). However, 6 Myr is a sufficiently long time, if compared to the expected duty cycle of such sources. In fact, IMBHs are expected to rapidly heat the surrounding gas and enhance the pressure, dramatically quenching the accretion rate on a time scale shorter than $6\times{}10^5$ yr (Krolik 2004). 

\section{IMBHs accreting molecular gas}
The aim of this paper is to constrain the number of IMBHs in HoII following the same technique adopted for the Milky Way by MFR. The basic idea is that, if a certain number of IMBHs are hosted in HoII and pass through dense gas regions, they should accrete this gas and possibly become observable as X-ray sources.
\begin{figure*}
\center{{
\epsfig{figure=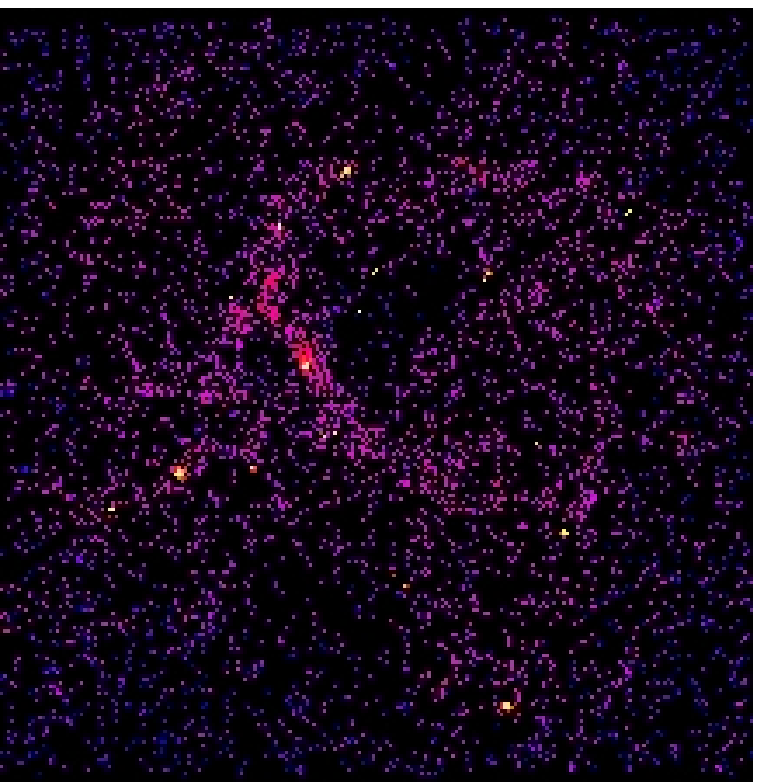,height=7cm}
\epsfig{figure=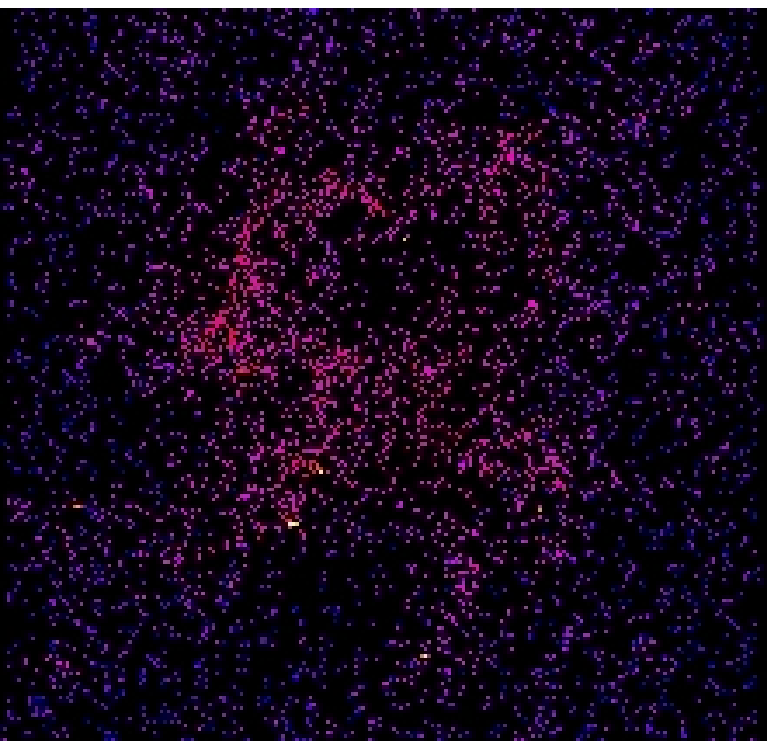,height=7cm}
}}
\caption{\label{fig:fig2}
Face-on projection of the central 200 pc of the simulated gas disc after 3 Myr. The color coding shows the z-averaged gas density. The scale, which is logarithmic, approximately goes from 0.1 (black; blue in the online version) to 10$^4$ (white; bright yellow in the online version) cm$^{-3}$. The bright yellow spots correspond to high-density regions hosting IMBHs. Left panel: run disc4\_{}100. Right panel: DMM4\_{}100. The number of accreting IMBHs is clearly higher for the case disc4\_{}100 than for DMM4\_{}100.
} 
\end{figure*}

Unfortunately, we do not know the details of the accretion process for IMBHs. As the simplest approximation (Mii \& Totani 2005; MFR), we can assume that these IMBHs accrete at the Bondi-Hoyle luminosity:
\begin{equation}\label{eq:bondihoyle}
L_X(\rho{}_g,\,{}v)=4\,{}\pi{}\eta{}\,{}c^2\,{}G^2\,{}m_\bullet{}^2\,{}\rho{}_g\,{}\tilde{v}^{-3},
\end{equation}
where $\eta{}$ is the radiative efficiency, $c$ the speed of light, $G$ the gravitational constant, $m_\bullet{}$ the IMBH mass, $\rho{}_g$ the density of the gas surrounding the IMBH. $\tilde{v}=(v^2+\sigma{}_{MC}^2+c_s^2)^{1/2}$, where $v$ is the relative velocity between the IMBH and the gas particles, $\sigma{}_{MC}$ and $c_s$ are the molecular cloud turbulent velocity and gas sound speed, respectively. We adopt $\sigma{}_{MC}=3.7$ km s$^{-1}$ (MFR and references therein), and we extract $v$ and $c_s$ directly from our simulations (see next section). $L_X$ in the above equation indicates the total X-ray luminosity. Properly speaking, the Bondi-Hoyle formula refers to the bolometric luminosity. However, in the case of ULXs the X-ray luminosity is much higher than the optical luminosity (Winter, Mushotzky \& Reynolds 2006), justifying our approximation.

The most slippery quantity in equation~(\ref{eq:bondihoyle}) is the efficiency $\eta{}$, which strongly depends on the details of the accretion process. Agol \& Kamionkowski (2002) show that density gradients in the molecular cloud are likely to imprint  a small amount of angular momentum on the accreting gas. This allows the formation of an accretion disc for most of IMBHs. However, this disc is not necessarily thin. It is more likely that IMBHs are surrounded by a thick disc, e.g. an ADAF disc (Narayan, Mahadevan \& Quataert 1998; Quataert \& Narayan 1999). The radiative efficiency of thick discs is quite lower than for thin discs.

In particular, the ADAF luminosity is $\propto{}\dot{M}^2$ (where $\dot{M}$ is the accretion rate) instead of $\propto{}\dot{M}$ as for the Bondi-Hoyle luminosity. However, for the systems we are considering
\begin{equation}
\frac{\dot{M}}{\dot{M}_{Edd}}\sim{}0.04\,{}\left(\frac{m_\bullet{}}{10^4\,{}M_\odot{}}\right)\,{}\left(\frac{n_{gas}}{100\textrm{ cm}^{-3}}\right)\,{}\left(\frac{\tilde{v}}{40 \textrm{ km s}^{-1}}\right),
\end{equation}
where $\dot{M}_{Edd}$ is the Eddington accretion rate and $n_{gas}$ the gas density. For the range $\dot{M}/\dot{M}_{Edd}\sim{}10^{-1}-10^{-2}$ it is reasonable to assume, as a lower limit, $\eta{}=10^{-3}$ (Narayan et al. 1998).

Then, in this paper we will assume  $\eta{}=10^{-3}$ as a conservative reference value. According to equation~(\ref{eq:bondihoyle}), our results can be easily rescaled for different values of $\eta{}$.

\subsection{Gas density and velocity}
The three terms in equation~(\ref{eq:bondihoyle}) that we can directly derive from our simulations are $\rho{}_g$, $v$ and $c_s$.

The high resolution of our simulations allows us to compute the gas density $\rho{}_g$  close to the IMBH accretion radius
\begin{equation}
r_{acc}=\frac{G\,{}m_\bullet{}}{v^2}\sim{}0.03\,{}\textrm{pc}\,{}\left(\frac{m_\bullet{}}{10^4\,{}M_\odot{}}\right)\,{}\left(\frac{v}{40\,{}\textrm{ km s}^{-1}}\right)^{-2}.
\end{equation}

In particular, in order to derive the density $\rho{}_g$ required in equation~(\ref{eq:bondihoyle}), we count the gas particles which fall within a certain neighborhood of each IMBH. The radius of the neighborhood has been chosen to be $r_g=5$ pc for molecular hydrogen and $r_g=100$ pc for atomic hydrogen. In fact, the adopted softening allows us to assume radii as small as $0.1-1$~pc; but we require each neighborhood to host a number of gas particles sufficient ($\gtrsim{}10$) to smear out numerical fluctuations. Then, $\rho{}_g$ has been derived as
\begin{equation}
\rho{}_g=\frac{{\rm N}_{g}\,{}m_{g}}{\frac{4}{3}\,{}\pi{}\,{}r_g^3},
\end{equation}
where N$_{g}$ is the number of gas particles in each  neighborhood of radius $r_g$ and $m_{g}$ is the mass of a gas particle ($m_{g}=2\times{}10^2M_\odot{}$ for molecular gas; see Section 3.3).

Because of the adopted polytropic equation~(\ref{eq:eq4}), the gas is allowed to become clumpy (Escala et al. 2005). Therefore, IMBHs particles attract nearby gas particles, rapidly increasing  their surrounding density. This is particularly evident for disc BHs and for a uniform disc of gas. In Fig.~\ref{fig:fig2}, the two runs disc4\_{}100 (left panel) and DMM4\_{}100 (right panel) are compared. The two panels of this figure show the face-on projected density of the gas in the central region of the simulated galaxy after 3 Myr. The bright spots represent the highest density clumps ($n_{gas}\gtrsim{}10^3$ cm$^{-3}$) in the simulation. Each of these spots hosts a central IMBH. The density of gas clumps surrounding the IMBHs in the left panel (where the IMBHs are simulated as a disc population) is significantly higher than in the right panel (where the IMBHs follow a DMM profile).

\begin{figure}
\center{{
\epsfig{figure=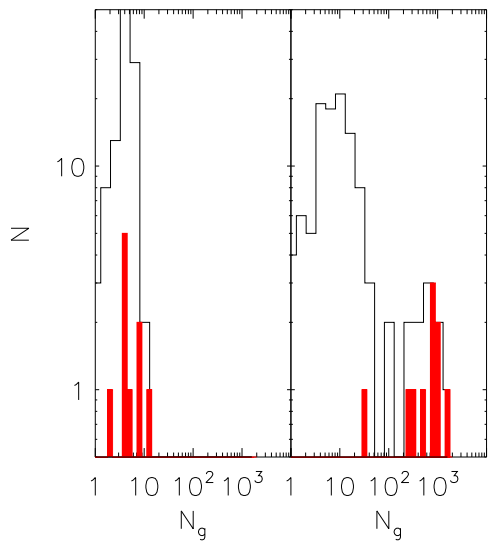,height=8cm}
}}
\caption{\label{fig:fig3} 
Simulation disc5\_{}10. Number of IMBHs (filled histograms) and gas particles (open histograms) which are surrounded by a number of gas particles N$_g$ (within a sphere of radius $r_g=5$ pc). The gas particles represented here are a sample of 100 particles randomly selected from the simulation.  Left panel: initial conditions. Right panel: after 6 Myr. After 6 Myr the number of gas particles N$_g$ surrounding the IMBHs has increased by a factor of $\sim{}100$.
}
\end{figure}
Because IMBHs attract nearby gas particles, the uniform gas disc becomes more and more clumpy and the density $\rho{}_g$ in the neighborhoods of the IMBHs increases (see Fig.~\ref{fig:fig3}), depending on the IMBH mass, position and velocity. This has a significant effect on the estimated luminosity (see next section for details).

Since we sampled  $\rho{}_g$ in neighborhoods of radius $r_g=5$ pc ($r_g=100$ pc for atomic gas) surrounding the IMBHs, a natural choice for deriving the relative velocity $v$  is to calculate the average relative velocity between the IMBH and the gas particles hosted in the same neighborhood.

Similarly, the sound speed of the gas, $c_s$, around each IMBH is calculated as the average sound speed of gas particles within the same neighborhood of radius $r_g$, by using the relation $c_s^2=2\,{}u_g$ (where $u_g$ is the average internal energy per unit mass of the gas particles within the considered neighborhood). The values of $c_s$ calculated with this method strongly depend on the density within the neighborhood, thus reflecting the feedback of the IMBH on the gas particles. In fact, the IMBH increases the gas density in its neighborhood and consequently heats the gas (because of the polytropic equation of state), quenching the accretion rate. For example, for disc IMBHs (the more efficient in attracting gas) we find $c_s\sim{}1-10$ km s$^{-1}$ (depending on $\rho_g$), quite higher than the value $c_s=0.3$ km s$^{-1}$ estimated for unperturbed Galactic molecular clouds (see MFR and references therein). Of course, this is not a realistic treatment of the accretion process; but it allows us to approximately account for the feedback from the IMBH.

\subsection{Comparison with observed X-ray sources}
We can now estimate the X-ray luminosity of the simulated IMBHs directly from these values of $\rho{}_g$, $v$ and $c_s$.
\begin{figure*}
\center{{
\epsfig{figure=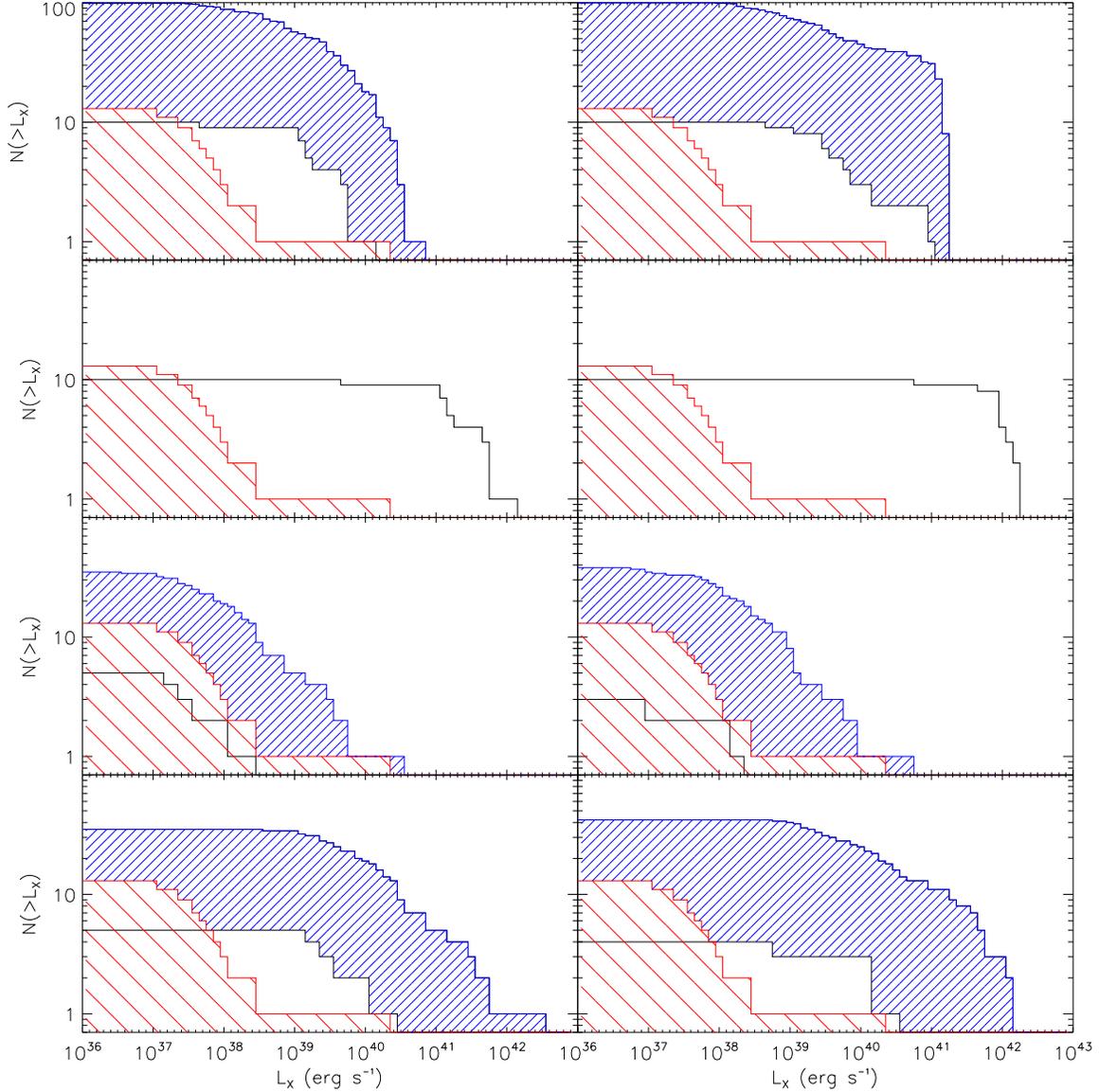,height=16cm}
}}
\caption{\label{fig:fig4} 
Cumulative distribution of observed and simulated X-ray sources as a function of the X-ray luminosity $L_X$.  The lightly hatched histogram reported in all the panels is the cumulative distribution of observed X-ray sources (Kerp et al. 2002; the luminosity of HoII X-1 is assumed to be in its brightest state, $L_X=2\times{}10^{40}$ erg s$^{-1}$, Dewangan et al. 2004). Open and heavily hatched histograms refer to simulations.
Left panels: initial conditions; right panels: after 6 Myr. 
First row (from top to bottom):  disc4\_{}10 (open histogram) and disc4\_{}100 (heavily hatched); second row: disc5\_{}10 (open); third row: DMM4\_{}100 (open) and DMM4\_{}1000 (heavily hatched); fourth row: DMM5\_{}100 (open) and  DMM5\_{}1000 (heavily hatched).} 
\end{figure*}

In Fig.~\ref{fig:fig4} we show the cumulative distribution of X-ray sources as a function of $L_X$ for the most significant cases and assuming $\eta{}=10^{-3}$.  The luminosity distributions can be easily rescaled for different values of $\eta{}$ according to equation~(\ref{eq:bondihoyle}). The left panels show the initial conditions, while the right panels represent the evolution after 6 Myr.

The IMBHs have a significantly different behaviour depending on their distribution (exponential or DMM). Disc IMBHs tend to attract gas particles and to increase their luminosity by a factor of $\sim{}$2-10 in 6 Myr, depending on their mass. 

On the contrary, most of the halo IMBHs do not attract efficiently gas particles. Halo IMBHs do not increase the density $\rho_g$ and maintain nearly the same luminosity during all the simulation.

This is probably due to the different velocity distributions. In fact, disc IMBHs are substantially corotating with the gas particles; whereas halo IMBHs have (on average) higher relative velocities and periodically leave the gaseous disc.

This result strongly affects any attempt to constrain the number of IMBHs.
In fact, if IMBHs are a disc population (as predicted, for example, by the runaway collapse scenario; Portegies Zwart \& McMillan 2002), the constraints on their number are very strong (see also Krolik 2004). Even 10 IMBHs of $10^4\,{}M_\odot{}$, corresponding to $\Omega{}_\bullet{}=10^{-4}\Omega{}_b$, would overproduce the number of X-ray sources in the range from $10^{38}$ to $10^{40}$ erg s$^{-1}$, unless their radiative efficiency is $\lesssim{}10^{-4}$ (i.e. a factor of 10  lower than the assumed $\eta{}=10^{-3}$). 


In the case of $10^5\,{}M_\odot{}$ IMBHs the constraints are even stronger: HoII cannot host 10 IMBHs with  $m_\bullet{}=10^5\,{}M_\odot{}$, unless their efficiency is $\eta{}\lesssim{}10^{-6}$. 

From these results, we  derive the following  upper limit for disc IMBHs:
\begin{equation}\label{eq:limdisc}   
\Omega{}_\bullet{}<10^{-5}\Omega{}_b\,{}\left(\frac{\eta{}}{10^{-3}}\right)^{-1}\,{}\left(\frac{m_\bullet{}}{10^4\,{}M_\odot{}}\right)^{-1}.
\end{equation}
This approximate formula is valid for $10^{4}-10^{5}M_\odot{}$ IMBHs. We made some checks for more massive BHs ($10^6-10^7M_\odot{}$) and we found that it is still applicable; but we cannot verify whether it holds also for lower masses. 
In equation~(\ref{eq:limdisc}) the scaling with $\eta{}^{-1}$ is entirely due to the fact that we adopted the Bondi-Hoyle formula (equation~\ref{eq:bondihoyle}) to calculate the luminosity. The scaling with $m_\bullet{}^{-1}$ is also due to the Bondi-Hoyle formula and to the fact that $\Omega{}_\bullet{}$ scales with the IMBH mass (see equation~\ref{eq:numBH}). 


The constraints on halo IMBHs are looser by a factor of $\sim{}10^3$. 1000 IMBHs with mass $m_\bullet{}=10^{4}\,{}M_\odot{}$ (i.e. $\Omega{}_\bullet{}=10^{-2}\Omega{}_b$) overproduce the number of X-ray sources if their efficiency is $\eta{}\gtrsim{}10^{-3}$. 
Similarly, 1000 halo IMBHs with mass $m_\bullet{}=10^5\,{}M_\odot{}$  exceed the number of X-ray sources in HoII for $\eta{}\gtrsim{}10^{-5}$. 

Thus, an upper limit valid for $10^{4}-10^{5}M_\odot{}$ halo IMBHs (distributed according to DMM) is:

\begin{equation}\label{eq:limDMM}  
\Omega{}_\bullet{}<10^{-2}\Omega{}_b\,{}\left(\frac{\eta{}}{10^{-3}}\right)^{-1}\,{}\left(\frac{m_\bullet{}}{10^4\,{}M_\odot{}}\right)^{-1},
\end{equation}
which, a part for the normalization, is the same as equation~(\ref{eq:limdisc}).
However, this constraint is more affected by statistical fluctuations, because of the small fraction of IMBHs ($\sim{}3-5$ per cent) which efficiently accrete with respect to the total number.

This upper limit is similar to the analogous constraint that MFR found for 10$^4\,{}M_\odot{}$ IMBHs distributed according to a DMM profile  in the Milky Way ($\Omega{}_\bullet{}<10^{-2}\Omega{}_b$ for $\eta{}=10^{-3}$). In this paper, the same upper limit has been derived on the basis of more robust simulations, as MFR used a simple Monte Carlo treatment for the gas instead of SPH simulations.

\begin{figure}
\center{{
\epsfig{figure=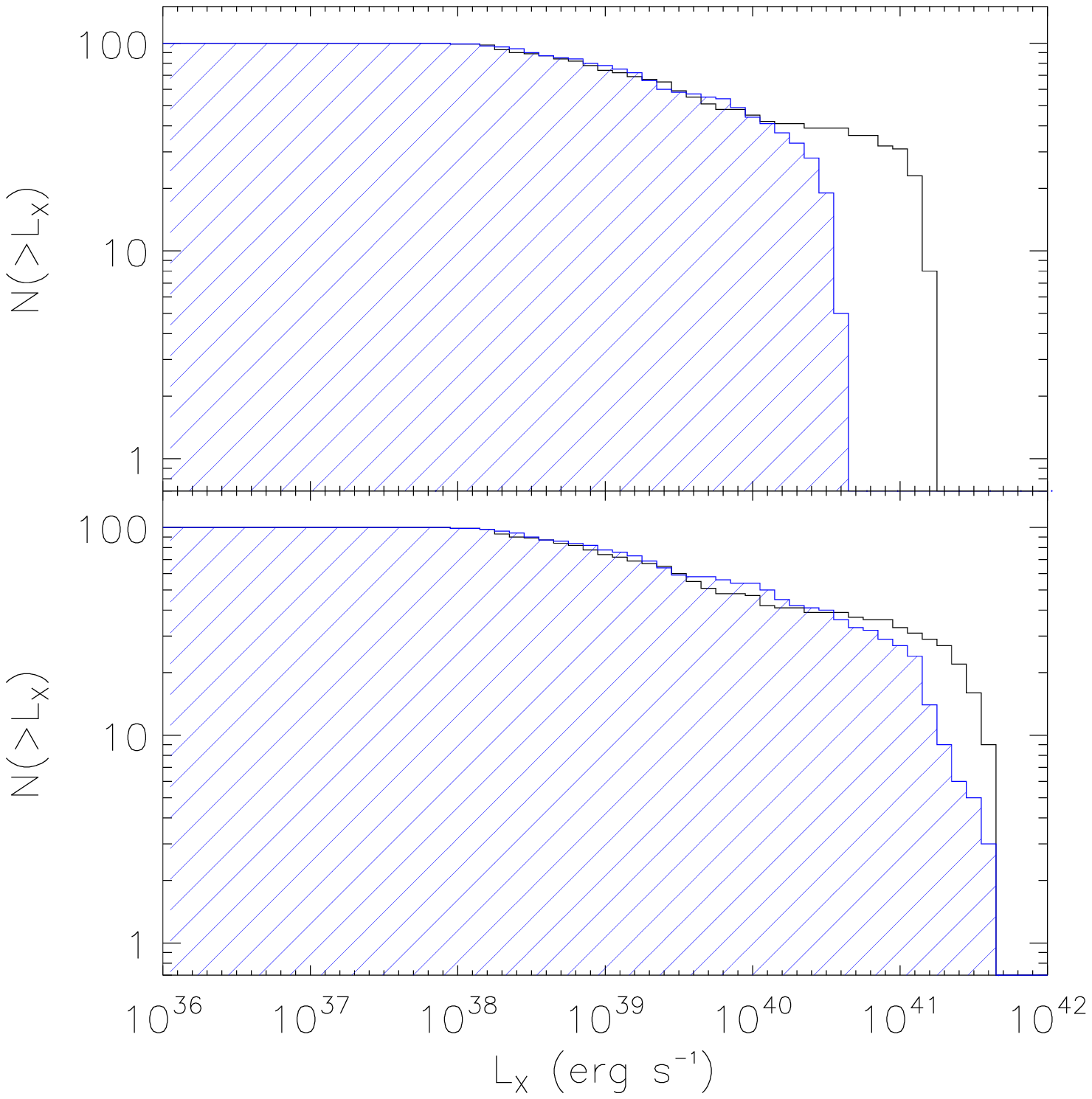,height=8cm}
}}
\caption{\label{fig:fig5} 
Cumulative distribution of X-ray sources as a function of the X-ray luminosity $L_X$, after 6 Myr, for the runs disc4\_{}100 (open histogram) and K\_{}disc4\_{}100 (hatched). Top panel: $c_s$ derived from the simulations. Bottom panel: $c_s=0.3$ km s$^{-1}$.
} 
\end{figure}

As we said, the efficiency of IMBHs to attract clumps of gas particles is important in determining the cumulative luminosity distribution. But the clumpiness of the simulated disc is function of the free parameter $\kappa{}$. In the simulations showed in Fig.~\ref{fig:fig4}  $\kappa{}=9\times{}10^{23}$ g$^{-2/3}$ cm$^4$ s$^{-2}$, which corresponds to a gas temperature $T_{gas}\sim{}60$ K (i.e. the average temperature of gas in molecular clouds). What happens if we change $\kappa{}$? Are our results sensitive to this parameter? In Fig.~\ref{fig:fig5} we compare the cumulative luminosity functions, after 6 Myr, of the runs disc4\_{}100 (where $\kappa{}=9\times{}10^{23}$ g$^{-2/3}$ cm$^4$ s$^{-2}$) and K\_{}disc4\_{}100 (where $\kappa{}=1.5\times{}10^{25}$ g$^{-2/3}$ cm$^4$ s$^{-2}$, corresponding to $T_{gas}\sim{}1000$ K). The two distributions are similar. The only significant difference is in the high-luminosity tail, because the warmer gas becomes less clumpy and its sound speed $c_s$ is higher (the average value of $c_s$ around the IMBHs is $\langle{}c_s\rangle{}\sim{}3$ km s$^{-1}$ for disc4\_{}100 and $\langle{}c_s\rangle{}\sim{}6.5$ km s$^{-1}$ for K\_{}disc4\_{}100). Then, $\kappa{}$ does not significantly influence our results, at least for a reasonable range of values. Of course, higher $\kappa{}$ imply larger Jeans masses, and then larger scale lengths of the clumps. However, these differences do not substantially affect the region close to the IMBH, where we sample the density $\rho{}_g$. 

In Fig.~\ref{fig:fig5} we can also note the effect of adopting for $c_s$ the values derived from the simulation (see Section 4.1). While most of the distribution is unchanged, the high-luminosity tail is strongly affected by adopting the simulated $c_s$ (upper panel) instead of the expected value (lower panel) for an unperturbed cloud (i.e. $c_s=0.3$ km s$^{-1}$). This is due to the fact that the highest density clouds around the IMBHs are also the hottest. In the case of the run disc4\_{}100 the luminosity distribution drops at $L_X\gtrsim{}1.8\times{}10^{41}$ erg s$^{-1}$, when the simulated $c_s$ is adopted (top panel); while, if the sound speed is assumed to be $c_s=0.3$ km s$^{-1}$, the brightest sources have $L_X\sim{}4.5\times{}10^{41}$ erg s$^{-1}$ (bottom panel).  The difference is even larger for the run K\_{}disc4\_{}100, where the gas has higher initial temperature ($T_{gas}=1000$ K): the high-luminosity tail reaches $L_X=4.5\times{}10^{41}$ erg s$^{-1}$ if $c_s=0.3$ km s$^{-1}$ (bottom panel), and stops at  $L_X=4\times{}10^{40}$ erg s$^{-1}$ if we derive $c_s$ from the internal energy (top panel).

\begin{figure}
\center{{
\epsfig{figure=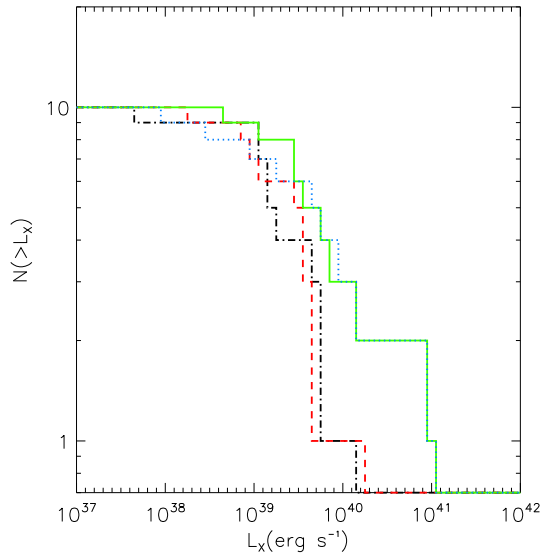,height=8cm}
}}
\caption{\label{fig:fig6} 
Cumulative distribution of X-ray sources as a function of the X-ray luminosity $L_X$ for the run disc4\_{}10 at the beginning of the simulation (dot-dashed line), after 2 Myr (dashed), after 6 Myr (solid)  and after 10 Myr (dotted). The luminosity rapidly increases in the first 6 Myr and then stabilizes.
} 
\end{figure}

As a sanity check of the stability of our results we also protracted some runs (disc4\_{}10, K\_{}disc4\_{}100, DMM4\_{}1000 and   HI\_{}DMM5\_{}100) for 10 Myr. We concluded that the situation at 10 Myr is not statistically different from the situation at 6 Myr, with respect to the inferred accretion rate of IMBHs, as most of the density evolution occurs in the first 5 Myr and the IMBH luminosity distribution remains quite stable in the next million years (Fig.~\ref{fig:fig6}). 

\section{Atomic hydrogen}      
MFR showed that in the Milky Way even IMBHs accreting atomic hydrogen can become observable as X-ray sources and are crucial to constrain the number of IMBHs. 
Thus, we considered atomic hydrogen also for HoII. Unfortunately, we cannot simulate a two-phase medium. Thus, in some runs we substituted molecular hydrogen with an exponential disc of gas. The total mass of gas as well as the scale length and scale height have been chosen according to the observational values (see Sections 2 and 3.1).

To calculate the luminosity of IMBHs accreting atomic gas, we adopt the same procedure described for molecular hydrogen (Section 4), except for the choice of the radius $r_g$. In fact, given the lower density of atomic hydrogen regions, we adopt $r_g=100$ pc, much larger than for molecular hydrogen.

\begin{figure}
\center{{
\epsfig{figure=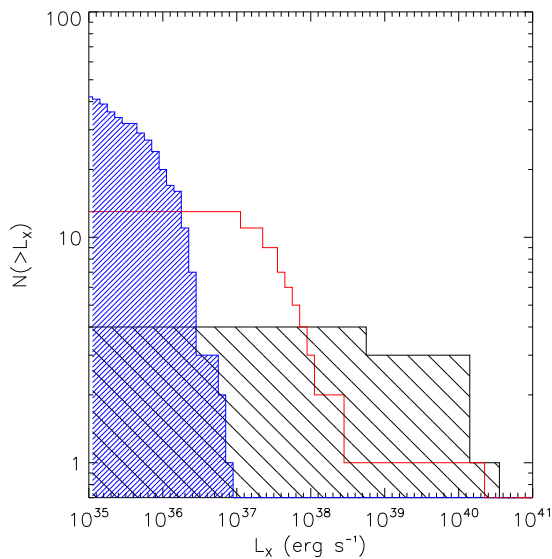,height=8cm}
}}
\caption{\label{fig:fig7} 
Cumulative distribution of X-ray sources as a function of the X-ray luminosity $L_X$, after 6 Myr, for the runs HI\_{}DMM5\_{}100 (heavily hatched histogram) and DMM5\_{}100 (lightly hatched). The open histogram shows the observations (Kerp et al. 2002; Dewangan et al. 2004).
} 
\end{figure}

To avoid contamination with IMBHs accreting molecular hydrogen, we do not consider those IMBHs that are inside the volume which should be occupied by the molecular hydrogen disc (i.e. within a scale length $R_{mol}=500$ pc and a scale height $z_{mol}$=60 pc). 

Fig.~\ref{fig:fig7} shows the results for the run HI\_{}DMM5\_{}100, i.e. the case where the contribution of atomic hydrogen has been found to be more important. It is apparent that IMBHs accreting atomic hydrogen (heavily hatched histogram) have luminosity $L_X<10^{37}$ erg s$^{^{-1}}$. Thus, these sources are below the observational threshold of Kerp et al. (2002), and are not relevant for our purposes.

This is not in contradiction with the results of MFR. In fact, also the luminosity distribution of IMBHs accreting atomic hydrogen in the Milky Way drops at $L_X=10^{37}$ erg s$^{-1}$. The difference between MFR and this paper consists merely in the data, which in the case of the Milky Way are complete down to 10$^{36}$ erg s$^{-1}$, rather than to 10$^{37}$ erg s$^{-1}$ as in HoII. In addition, most of Galactic sources have been identified as low mass X-ray binaries, high mass X-ray binaries and other classes of known sources, excluding the possibility of being IMBHs.

\section{Summary}     
In this paper we derived upper limits on the density of IMBHs in HoII, by assuming that they can accrete dense gas and emit as X-ray sources. 
In particular, we ran SPH N-body simulations of a disc dwarf galaxy having the main observational properties of  HoII and hosting a halo or disc population of IMBHs. We compared the luminosity distribution derived by these simulations with the observational data by Kerp et al. (2002). 
A similar technique has been previously applied to the Milky Way by MFR.

The target galaxy of the current paper, HoII, was chosen for its rich content in gas and X-ray sources, as well as for its small mass (if compared to the Milky Way), which enables us to reach very high mass and spatial resolution if compared to MFR.
The improvements with respect to MFR are not only in the resolution, but also in the treatment of the gas (this paper is based on SPH simulations, whereas MFR used a Monte Carlo approach) and in the presence of a 'living' halo (MFR adopted a rigid halo).

MFR assumed that IMBHs, if exist, are a (more or less concentrated) halo population. This might be the case if IMBHs are the  remnants of massive population III stars formed in minihalos. But other formation mechanisms, such as the runaway collapse in young clusters (Portegies Zwart \& McMillan 2002), suggest that IMBHs can also be a disc population.  IMBHs could reside in discs even if they are remnants of population III stars formed at relatively low redshift (Krolik 2004; Jimenez \& Haiman 2006). Thus, in this paper we consider both the case where IMBHs are a halo population (following a DMM profile) and the case where IMBHs are a concentrated disc population.

If IMBHs are a disc population, we find an upper limit $\Omega{}_\bullet{}<10^{-5}\Omega{}_b\,{}\eta{}_{3}^{-1}\,{}m_4^{-1}$ (where $\eta{}_{3}=\eta{}/10^{-3}$ and $m_4=m_\bullet{}/10^4\,{}M_\odot{}$, equation~\ref{eq:limdisc}).

In the case IMBHs follow a DMM profile, the upper limit can be written as $\Omega{}_\bullet{}<10^{-2}\Omega{}_b\,{}\eta{}_{3}^{-1}\,{}m_4^{-1}$ (equation~\ref{eq:limDMM}), similar to the upper limit found by MFR for the Milky Way.

Thus, limits on disc IMBHs are a factor of 1000 stronger than limits on halo IMBHs. This implies that only 1 IMBH with mass $10^4\,{}M_\odot{}$ can be hosted in the disc of HoII, unless the radiative efficiency is lower than $10^{-3}$. Such limits have important consequences on models of IMBH formation such as the runaway collapse scenario.

In summary, the results presented here confirm the results by MFR, and strengthen the upper limits derived from the Milky Way. Unfortunately, this method suffers of a number of uncertainties. First of all, the radiative efficiency $\eta{}$ was left as a free parameter (even if we adopted $\eta{}=10^{-3}$ as a reference value for the figures). In addition, our method does not account for the effective duty cycle of such IMBHs. As suggested by Krolik (2004), the duration of the high-luminosity state of these sources must be very short (less than 1~Myr), because the IMBH rapidly heats the surrounding gas, quenching the accretion rate. This effect cannot be considered in our simulations, and should be further investigated.


\section*{Acknowledgements}
The author thanks E. Ripamonti, L.~Mayer, A. Ferrara, L.~Tornatore, S. Sigurdsson and M. Eracleous for useful
discussions, the anonymous referee for a critical reading of the manuscript, and acknowledges the technical support from the staff at Cilea and the system managers of the Kapteyn Astronomical Institute of Groningen.

\onecolumn
\appendix

\end{document}